\documentstyle[psfrag,preprint,graphicx,tighten,floats,aps]{revtex}



\def\A{{\cal A}}
\def\Ab{\bar{\A}}
\def\C{{\cal C}}
\def\G{{\cal G}}
\def\Gb{\bar{\cal G}}
\def\H{{\cal H}}
\def\HF{{\H}_F}
\def\M{{\cal M}}
\def\N{{\cal N}}

\def\S{{\cal S}}

\def\r{{(r)}}
\def\hb{\bar{A}}
\def\n{\vec{n}}
\def\cyl{{\rm Cyl }}
\def\SU{{\rm SU}}
\def\U{{\rm U}}
\def\lp{\ell_{\rm Pl}}

\def\be{\begin{equation}}
\def\ee{\end{equation}}
\def\ba{\begin{eqnarray}}
\def\ea{\end{eqnarray}}

\begin{document}
\title{Relation between polymer and Fock excitations}
\author{Abhay\ Ashtekar${}^{1,3}$, Jerzy Lewandowski${}^{2,1,3}$} 
\address{1. Physics Department, 104 Davey, Penn State, University
Park, PA 16802, USA\\ 2. Institute of Theoretical Physics, University
of Warsaw, ul. Ho\.{z}a 69, 00-681 Warsaw, Poland\\ 3. Max Planck 
Institut f\"ur Gravitationsphysik, Albert Einstein Institut, 14476
Golm, Germany}

\maketitle

\begin{abstract}

To bridge the gap between background independent, non-perturbative
quantum gravity and low energy physics described by perturbative field
theory in Minkowski space-time, Minkowskian Fock states are located,
analyzed and used in the background independent framework.  This
approach to the analysis of semi-classical issues is motivated by recent
results of Varadarajan. As in that work, we use the simpler $U(1)$
example to illustrate our constructions but, in contrast to that work,
formulate the theory in such a way that it can be extended to full
general relativity.
\end{abstract}
\bigskip

\textit{Motivation}
\medskip

A detailed theory of quantum geometry was systematically developed in
the mid-nineties [1-17].  Over the last two years, it was successfully
applied to address some of the long-standing challenges of quantum
gravity. These include: a statistical mechanical derivation of black
hole entropy \cite{abck}; a resolution of the big-bang singularity in
quantum cosmology \cite{mb}; and, a ``finiteness result'' in the path
integral (or spin-foam) approach to quantum general relativity
\cite{ap}. These results make a crucial use of the fundamental
discreteness, predicted by the background independent quantum
geometry. But this very discreteness has made it difficult to relate
the underlying `polymer excitations' of quantum geometry to the
Fock states normally used in low energy physics. Candidate
semi-classical states of quantum geometry \textit{have been}
introduced by a number of authors (for early ideas, see
\cite{ars1,ma,vz,cr} and, for a detailed framework, \cite{toh}).
However, there is no obvious relation between them and the Fock
coherent states, normally used to analyze semi-classical issues in low
energy physics.  As a result, detailed answers to two key questions
are yet to emerge: Can the background-independent, non-perturbative
theory reproduce the familiar low-energy physics on, say, suitable
coarse graining?  and, Can one pin-point where and why perturbation
theory fails?

More precisely the situation is as follows. Since the approach begins
with general relativity coupled to matter (or, supergravity) and
carries out a (canonical or path integral) quantization following
standard procedures, at the simplest level it is clear that the
quantum theory would have the correct classical limit.  Similarly, the
fact that the fundamental discreteness arises at the Planck scale
\cite{ars1} strongly suggests that while a coarse graining at, say, a
TeV scale should reproduce the continuum physics, the ultraviolet
behavior of the non-perturbative and perturbative theories would be
very different.  However, these can only be regarded as general
indications.  What is lacking is a systematic investigation leading to
detailed answers to the two question.

Efforts at facing this challenge squarely have begun recently.  A
point of departure is provided by the recent work of Varadarajan
\cite{mv}. The purpose of this letter is outline the general approach
and summarize some of the key results. Detailed proofs and further
results will appear elsewhere \cite{afl}.
\bigskip

\textit{The tension}
\medskip

In fully non-perturbative approaches to quantum gravity,
semi-classical issues are conceptually difficult because there is no
background space-time to begin with.  Furthermore, the basic
mathematical structures in these approaches are very different from
those used in more familiar perturbative treatments.  For example, in
the quantum geometry framework, fundamental excitations are one
dimensional, polymer-like [1-5]; a convenient basis of states is
provided by spin-networks \cite{rp,rs2,jb2,jb3}; and, eigenvalues of
the basic geometric operators defining triads \cite{al4}, areas and
volumes \cite{rs3,al4,al5} are discrete.  By contrast, in the Fock
framework, the fundamental excitations are 3-dimensional, wavy; the
convenient basis is labelled by the number $n$ of gravitons, their
momenta and helicities; and, all geometric operators have continuous
spectra.  The challenge is to bridge the gap between these apparently
disparate frameworks.  In most of this letter, for concreteness and
simplicity, we will consider an $U(1)$ model which captures the key
features of the problem at hand.  However, our resolution of the
tension between the two frameworks is well suited for generalization
to full quantum gravity.

Let us begin by using the $U(1)$ example to bring out the tension in 
mathematical terms.  Fix a space-like 3-plane $M$ in Minkowski space.  To 
construct the $U(1)$-analog of quantum geometry [1-5], let us begin by 
introducing the quantum configuration space $\Ab$ for the $U(1)$ theory.  
An element $\hb$ of $\Ab$, called a generalized connection, associates with 
every oriented, analytical edge $e$ in $M$ a holonomy, i.e., an element of 
$U(1)$, such that: i) $\hb(e^{-1}) = [\hb(e)]^{-1}$; and, ii) if $e_1$, 
$e_2$ and $e_3:= e_1\cdot e_2 $ are all analytic edges, then $\hb(e_3) = 
\hb(e_1)\, \hb(e_2)$.%
\footnote{It is often convenient to quotient $\Ab$ by the action of
generalized gauge group $\Gb$ consisting of (arbitrarily
discontinuous) maps from $M$ to $U(1)$. The resulting space $\Ab/\Gb$
is also the Gel'fand spectrum of the obvious holonomy algebra
constructed from smooth connections on $M$ \cite{ai1}.}
The space $\Ab$ carries a natural, diffeomorphism invariant measure
$\mu_o$, induced by the Haar measure on $U(1)$. Following the quantum
geometry strategy \cite{al1,jb1,al2,mm}, let us choose $\H_o :=
L^2(\Ab, \mu_o)$ as the Hilbert space of states. There are two
families of operators defined naturally on $\H_o$. The first,
$\hat{h}_\gamma$, are \textit{holonomy operators},  labelled by
closed loops $\gamma$. They are unitary and operate by
multiplication. The second family, $\hat{E}_S$, is labelled by
2-surfaces $S$ without boundary. These are self-adjoint operators,
representing the flux of the electric field through $S$. Thus, in this
framework, the connection is smeared along 1-dimensional loops and the
electric field along 2-surfaces.

Eigenvalues of the electric flux operators are integers; \textit{in this 
representation, the electric flux is quantized}.  The eigenstates, called 
\textit{flux networks}, provide a useful orthonormal basis in $\H_o$.  
Given any graph $\alpha$ with $N$ oriented edges, and an assignment of an 
integer to each edge, we can define a flux network function $\N_{\alpha, 
\n}$ on $\Ab$ via
\be \label{fn}
\N_{\alpha, \n}\, (\hb) := [\hb(e_1)]^{n_1} \ldots [\hb(e_N)]^{n_N} . 
\ee
Given any 2-surface $S$, the action of $\hat{E}_S$ on this state is 
given by \cite{al4}:
\be \hat{E}_S \,\N_{\alpha, \n} = \left(\frac{\hbar}{2}\, \sum_I
\, k_I \, \epsilon_I\, n_I\right)\, \N_{\alpha, \n} \ee
where the sum is over edges of $\alpha$ which intersect $S$; $k_I =
\pm 1$ depending on whether the $I$th edge lies above or below $S$ and
$\epsilon_I = \pm 1$ depending on whether the $I$th edge is oriented
to leave $S$ or arrive at $S$. Thus, the labelled edges of $\alpha$
can be regarded as flux lines of the electric field operator, the
integer $n_I$ denoting the number of `fluxons' along the $I$th edge.
Being 1-dimensional , these excitations are said to be `polymer-like'.

We will refer to finite linear combinations of flux networks as
cylinder functions and denote by $\cyl$ the dense subspace of $\H_o$
they span.%
\footnote{This is a slightly different usage of the term from 
\cite{al1,jb1,al2,mm,al3} but more convenient for our purposes here.}
The algebraic dual of $\cyl$ will be denoted by $\cyl^\star$.  We thus have 
a triplet, 
\be \label{triple}
\cyl \subset \H_o \subset \cyl^\star\, ,
\ee
which is analogous to the Gel'fand triplet, often used in quantum
mechanics \cite{gt}. Quantum geometry has analogous structures: There,
spin networks are eigenstates of geometric operators, the kinematical
Hilbert space $\H_o$ enables one to obtain well-defined operators
representing quantum constraints, and physical states of quantum gravity
--i.e., solutions to the constraints-- belong to $\cyl^\star$. Note
that, as in quantum geometry, the entire framework --states as well as
the operators-- is diffeomorphism invariant; in particular, 
the  space-time metric is not used anywhere.

The Fock framework, on the other hand, makes a heavy use of the
space-time metric.  For our purposes, it is more convenient to use the
Schr\"{o}dinger representation. The Hilbert space $\HF$ is then given
by $\HF = L^2(\S^\star, d\mu_F)$, where $\S^\star$ is the space of
tempered distributions on $M$ and $\mu_F$ is the Gaussian measure
(defining the usual Fock space of photons). The basic operators
are now the smeared connection and electric fields, $\hat{A}(f)$ and
$\hat{E}(g)$. The connection and the electric field are both transverse 
operator-valued distributions and require a \textit{3-dimensional} smearing.  
Their spectra are continuous.

The challenge is to relate the two descriptions which appear to have
little in common. The natural `homes' for these quantum theories are
the quantum configuration spaces, $\Ab$ and $\S^\star$. They are quite
different spaces; while they admit a non-trivial intersection, neither
space is contained in the other. Therefore, at first sight it seems
difficult to even begin a comparison.
\bigskip\vfill\break

\textit{A Resolution}
\medskip

A primary source of the tension between the two frameworks lies in the
fact that while holonomies play a central role in the definition of
generalized connections in $\Ab$, the holonomy of a general element of
$\S^\star$ fails to be well defined because tempered distributions
have to be smeared in \textit{three} dimensions, while the holonomy
provides smearing along only one dimension \cite{ai2}.  To overcome this 
incompatibility, let us define a `contraction map' $\C_r$ on $\S^\star$.  
The Fourier transform of a tempered distribution is again a tempered 
distribution and it is easier to define the action of $\C_r$ in the 
momentum space:
\be
\C_r: \S^\star \mapsto \S^\star;\quad\quad  \C_r (A_a(\vec{k})) =
e^{ -\frac{1}{2}k^2r^2} \, A_a(\vec{k})\, =:  A_a^{(r)}(\vec{k}) 
\ee
where $r$ is any fixed positive real number. Let us denote by
$A_a^{(r)}(x)$ the Fourier transform of $A_a^{(r)}(k)$.  A simple
calculation shows that, given any smooth loop $\gamma$, the holonomy
of the connection $A_a^{(r)}(x)$ around $\gamma$ is
well-defined. Thus, $\C_r$ \textit{tames} tempered distribution
sufficiently to make their line-integrals well-defined.  Therefore, 
intuitively, one would expect that every $A_a^{\r}(x)$ defines an element 
of $\Ab$.  Using the characterization of $\Ab$ in terms of the so-called 
`hoop group' \cite{al1}, one can show that this expectation is correct.  
Thus, $\C_r$ is a 1-1, onto mapping from $\S^\star$ to $\S^\star \cap 
\Ab$.

We can now push forward the Fock-measure $\mu_F$ on $\S^\star$ to
obtain a measure $\mu_{F}^\r$ on $\Ab$.  Consequently, Fock states can
now be represented as functions on $\Ab$ which are square-integrable
with respect to $\mu_{F}^\r$. Of course, the actions of the Fock
operators $\hat{A}(f)$ and $\hat{E}(f)$ on this representation are
more involved \cite{mv}, obtained from the standard actions via
pull-backs and push-forwards defined by $\C_r$. This complication is
inevitable; if one wishes to retain the standard action, it is
impossible to have well-defined holonomy operators in the Fock
representation \cite{ai2}.  Nonetheless, for \textit{each} $r>0$,
$L^2(\Ab, d\mu_F^\r)$ provides a representation of the usual operator
algebra which is unitarily equivalent to the standard Fock
representation.

The conceptual simplification brought about by this step is that $\Ab$
\textit{can serve as a `common home' for both the `polymer' and the
Fock descriptions}. We can now meaningfully ask for the relation
between them.
\bigskip

\textit{The relation between measures}
\medskip

Since both $\mu_o$ and $\mu_F^\r$ are measures on the same space
$\Ab$, we can now ask for the relation between them, and, more
generally, investigate the relation between Fock states and elements
of the triplet $\cyl \subset \H_o \subset \cyl^\star$ in the polymer
description.
 
Given a graph $\alpha$, there is an associated quantum configuration
space $\A_\alpha$, isomorphic to $[U(1)]^N$, where $N$ is the number
of edges of $\alpha$.  $\A_\alpha$ is obtained simply by restricting
the action of generalized connections in $\Ab$ to the edges of the
graph $\alpha$ and can be regarded as the configuration space of an
$U(1)$ lattice gauge theory based on $\alpha$.  The Haar measure on
$U(1)$ provides a natural measure $\mu^o_\alpha$ on $\A_\alpha$ and
the flux network states $\N_{\alpha,\n}$ provide an orthonormal basis
for the Hilbert space $H^o_\alpha := L^2(\A_\alpha, d\mu^o_\alpha)$.
$\Ab, \mu_o$ and $\H_o$ can all be obtained as projective limits of
$\A_\alpha, \mu^o_\alpha$ and $H^o_\alpha$ in which one considers
larger and larger graphs $\alpha$ \cite{jb1,al2,mm,al3}.

Now, \textit{every} measure on $\Ab$ arises as a consistent
family of measures on $\A_\alpha$ associated with graphs $\alpha$
\cite{jb1,al2,mm,al3}.  This is in particular true of the measure
$\mu_F^\r$. Therefore, to understand the relation between Fock and
polymer excitations, we can compare the measures $\mu^o_\alpha$ and
$\mu_\alpha^{\r}$ on $\A_\alpha$, corresponding to the measures
$\mu_o$ and $\mu_F^\r$ on $\Ab$. A simple calculation shows that, for
any graph $\alpha$, the measure $\mu_\alpha^{\r}$ is absolutely
continuous with respect to $\mu^o_\alpha$. 

To exhibit the function relating them, we first recall \cite{ars2} the
notion of the $r$-form factor associated with any oriented edge
$e$. The form-factor $F^a_e(\vec{x})$ of $e$ is given by
\be
F^a_e(\vec{x}) := \int_e ds\, \dot{e}^a(s)\,\, 
\delta^3(\vec{x}, \vec{e}(s))
\ee
so that the holonomy of any connection $A^\r$ along $e$ is given by
$\int d^3x A_a^\r (\vec{x}) F^a_e (\vec{x})$.  (Note that $F^a_e(\vec{x})$
is insensitive to orientation-preserving re-parameterizations $s \rightarrow
 s'$ of $e$.) The `tamed' r-form factor is defined as:
$$ F^a_{e,r}(\vec{x}) := {1\over (2\pi)^{\frac{3}{2}}} \,\,
\int_e ds\,  \dot{e}^a(s)\,\, 
\frac{e^{ -\frac{|\vec{x} -\vec{e}(s)|^2}{2r^2}}}{r^3} $$
Denote by $F^a_e(\vec{k})$ the Fourier transform of
$F^a_e(\vec{x})$. Then the Fourier transform of the `tamed' r-form
factor is simply:
$$
F^a_{e,r}(\vec{k}) :=  e^{-\frac{1}{2}k^2r^2} \,\, F^a_e(\vec{k})\, 
$$

In terms of these r-Form factors, the two measures are related by
\be \label{rel}
d\mu_\alpha^{\r} =  \left[\sum_{\n} e^{ - \frac{\hbar}{4}\, 
\sum_{I,J} G_{IJ}\, n_I n_J} \,\, 
\N_{\alpha,\n}(\hb_\alpha)\right] \,\, d\mu_\alpha^o 
\ee
where $I,J$ range from $1$ to $N$ (the number of edges of $\alpha$), 
and 
\be
G_{IJ} = \int \frac{d^3k}{|k|} 
\,\, \bar{F}_{e_I,r}\,\cdot\, F_{e_J,r}\,\, ,
\ee
the $\cdot$ denoting the contraction of the vector indices of
form-factors with the natural Euclidean metric on $M$.  The measure
$\mu_\alpha^o$ knows nothing about space-time geometry while the
measure $\mu_\alpha^\r$ does. This information (as well as our `taming
procedure') is neatly coded in the $N\times N$ matrix $G_{IJ}$ on the
space of edges of $\alpha$. One can show that the infinite sum in the
square brackets on the right side of (\ref{rel}) converges to a
continuous (non-constant) positive definite function $\varphi_\alpha$
on (the compact space) $\A_\alpha$. Thus, given any graph, the two
measures are absolutely continuous with respect to one another and
both are faithful.  The second property implies that the measure
$\mu_F^\r$ is also faithful\cite{al2}. However, measures
$\mu_\alpha^\r$ obtained by varying the graph $\alpha$ can not be
pushforwards of a measure obtained by multiplying $d\mu_0$ by a
mesurable function $\varphi$ on $\Ab$ because there is no function on
$\Ab$ whose pull-backs (under natural projections from $\Ab$ to
$\A_\alpha$) equal the above mentioned functions $\varphi_\alpha$ on
$\A_\alpha$ for all graphs $\alpha$.
\footnote{After this work was completed, M. Varadarajan pointed out to
us a recent paper by J.M. Velhinho \cite{jv} in which measure
theoretic issues concerning $\mu_o$ and $\mu_F^\r$ are discussed in
detail.}

Finally, it is interesting to note that the Fock measure $\mu_F^\r$
can be obtained by taking a projective limit of measures
$\mu_\alpha^\r$ associated with (floating) lattices $\alpha$.  This is
an interesting and, for our purposes, crucial alternative to the
conventional procedure in which one begins with a fixed lattice and
lets the lattice spacing go to zero.

\bigskip

\textit{Fock states as elements of $\cyl^\star$}
\medskip

Each regular measure on $\Ab$ naturally gives rise to a representation
of the holonomy $C^\star$ algebra \cite{ai1,al1}.  Since the measures
$\mu_o$ and $\mu_F^\r$ are inequivalent, so are the two
representations.  Therefore, Fock states can not be realized as
elements of $\H_o$. In particular, there is no state in $\H_o$ in
which the expectation values of all holonomy operators equal those in
the vacuum state in $\H_F^\r$. Nonetheless, as shown below, every
coherent state (and hence every state) in $\H_F^\r$ can be naturally
regarded as an element of $\cyl^\star$. In full quantum gravity,
solutions to constraints naturally lie in $\cyl^*$. The fact that Fock
states used in the low energy, perturbative analyses also share this
`home' will facilitate the comparison between non-perturbative and
perturbative treatments.

In \cite{mv}, Varadarajan imposed the `poincar\'e invariance
condition' to single out, among elements of $\cyl^\star$, the vacuum
state of $\H_F^\r$. His calculation was tailored to the framework
developed in \cite{ars2} and thus used closed loops.  For our
purposes, it is more convenient to use flux network states. Note first
that since every element of $\cyl$ can be expressed as a finite linear
combination of flux network states, there is a natural basis in
$\cyl^\star$ consisting of elements of the type
$<\!\N_{\alpha,\n}\!\mid$ which maps the flux network function
$\mid\!\N_{\alpha,\n}\! >$ to one and every other flux network
function to zero \cite{almmt}. In terms of this basis, the element
$<\!  V_F^\r\!\mid$ of $\cyl^\star$ representing the vacuum in
$\H_F^\r$ can be written as follows:
\be \label{vac}
<\!  V_F^\r\!\mid \,\, = \,\,  \sum_{\alpha,\n}\, 
\left[ e^{ -\frac{\hbar}{2} \sum_{IJ} G_{IJ}n_I n_J} \right]\,
<\!\N_{\alpha,\n}\!\mid \, .
\ee
Although the sum is over an uncountable set, while acting on any
element of $\cyl$, only a finite number of terms are non-zero, whence
the action is well-defined.%
\footnote{The exponents on the right sides of (\ref{rel}) and
(\ref{vac}) are closely related but differ by a factor of 2:
while the expression in the square brackets in (\ref{rel}) is
the vacuum expectation value of the operator $\hat{\N}_{\alpha, \n}$
(which acts by multiplication), that in (\ref{vac}) is the action of 
the element $<\!  V_F^\r\!\mid$ of $\cyl^\star$ on the cylindrical
function $\N_{\alpha, \n}$.}

Note some interesting features of this construction.\\ i)
${\cyl}^\star$ and the basis $<\!\N_{\alpha,\n}\!\mid$ in it are
constructed in a diffeomorphism invariant fashion; these structures
know nothing about Minkowski geometry.  How is it then that we can
locate Minkowskian Fock states as elements of $\cyl^\star$?  The
information about space-time geometry gets fed in to the expression
(\ref{vac}) through the coefficients in the linear combination of
these basis vectors.\\ 
ii) These numerical coefficients have a non-local spatial dependence;
there is an `interaction' between all different edges of the graph
$\alpha$ with each other.  These non-local correlations are
characteristic of Fock states and have physical consequences.  In
particular, this non-locality is responsible for the fact that the
Fock vacuum represents the `Coulomb phase' of the $U(1)$ theory in
which Wilson and (their dual) 't Hooft loops both go as exponentials
of the length of the loop rather than area.  In the polymer
representation, one \textit{can} define semi-classical states without
such non-local correlations \cite{toh}.  But such states typically
belong to the `Higgs phase' (`dual' of the confined phase) in which
the Wilson loop goes as the exponential of the length but the 't Hooft
loop goes as the exponential of the area.  \\
iii) Since $\cyl^\star$ is the algebraic dual on $\cyl$, the
holonomy and electric flux operators $\hat{h}_\gamma$ and $\hat{E}_S$
on $\cyl$ have a natural, well-defined action on $\cyl^\star$ by
duality.  One can show that the subspace of $\cyl^\star$ obtained by
acting repeatedly by $\hat{h}_\gamma$ on $<\!  V_F^\r\!\mid$ is the
embedding of a dense subspace of $\H_F^\r$ in $\cyl^\star$.  In this
sense, all Minkowskian, photon Fock states are realized as elements of
$\cyl^\star$.  However, $\cyl^\star$ is \textit{ very large}: not only
does it contain such perturbative states on other space-time
geometries but it also contains states (such as the basis used in
(\ref{vac})) which lie \textit{entirely outside the semi-classical
regime}.  In the context of full quantum gravity, such states in
$\cyl^\star$ would represent quantum excitations of the geometry and
the $U(1)$ gauge field which can not be described in space-time
terms.\\ 
iv) It is because the \textit{ photon Fock states span a `very
small' sub-space of} $\cyl^\star$ that the Fock representation fails to
capture flux quantization ---i.e., fundamental discreteness--- of the
`polymer' representation.  For, while the basis
$<\!\N_{\alpha,\n}\!\mid$ in $\cyl^\star$ provides infinitely many
eigenvectors with discrete eigenvalues of the electric flux operators
$\hat{E}_S$, none of them belong to the image of $\H_F^\r$ in
$\cyl^\star$.  Indeed, although the operators $\hat{E}_S$ have a
well-defined action on $\cyl^\star$, \textit{they fail to leave the
image of $\H_F^\r$ in $\cyl^\star$ invariant}.  From the `polymer'
perspective this is why the flux operators fail to be well-defined in
the Fock space.

Finally, we can represent any coherent state as an element of
$\cyl^\star$.  Using the fact that coherent states are eigenstates of
annihilation operators, one can extend the above procedure, used for
the vacuum state, to represent a general coherent state, peaked at a
smooth transverse connection $A_o$ and electric field $E_o$, as an
element $<\!  \Psi^\r_{(A_o,E_o)}\!\mid$ of $\cyl^\star$:
\be \label{coh}
<\! \Psi_{(A_o,E_o)}^\r\!\mid \,\, =\,\,  \sum_{\alpha,\n}\, 
\left[ e^{ i\hbar\, \sum_I\, n_I \int {d^3k}\, 
\bar{F}_{I,r}(k)\cdot f(k)}
\,\, e^{ -\frac{\hbar}{2} \sum_{IJ} G_{IJ}\, n_I n_J} \right]\,
<\!\N_{\alpha,\n}\!\mid \, ,
\ee
where, $\hbar f(k) = A_o(k) - (i/ \mid\! k \! \mid)  E_o(k)$.
Again the information about all background fields ---the Minkowskian
geometry as well as the electromagnetic field $(A_o, E_o)$ which
approximates the semi-classical state--- is coded in the coefficients
of the linear combination.  As a result, although the space
$\cyl^\star$ and the basis $<\!\N_{\alpha,\n}\!\mid $ is constructed
in a non-perturbative, diffeomorphism invariant fashion, individual
elements of $\cyl^\star$ can still carry information about background
fields, needed in Minkowskian, low energy perturbation theory.

\textsl{Remark}: The coefficients of expansions in Eqs (\ref{vac}) and
(\ref{coh}) closely resemble the expressions of the vacuum and
coherent states obtained in \cite{loop} in the \textit{loop}
representation. However, there are conceptual as well as technical
differences.  On the conceptual side, the focus of that work was on
understanding the loop representation, in particular the probabilities
of occurrence of specific types of loops in these states, while in
this work we work primarily with the space $\Ab$ of generalized
connections and our goal is to represent Fock states as elements of
$\cyl^\star$. At the technical level, while the ultra-violet cut-off
used in \cite{loop} does make the loop states well-defined, it is not
obvious that the result is unitarily equivalent to the standard Fock
representation. As discussed above, our $r$-taming does yield an unitarily
equivalent representation.
\bigskip\vfill\break

\textit{Shadow states}
\medskip

At this stage of development of quantum geometry, $\cyl^\star$ does
not have a natural inner product with respect to which both the
eigenstates $ <\!\N_{\alpha,\n}\!\mid $ of the electric flux operators
and the (images of the) Fock states are normalizable.  One obvious
strategy is to simply ignore those elements of $\cyl^\star$ which do
not represent Fock states and use the standard Fock norm (provided by
$\mu_{F}^{\r}$) on those which do.  But then the non-perturbative,
`polymer' perspective would be entirely lost and one would just be
reproducing the Fock representation in an unnecessarily complicated
fashion.  What we need is a new structure which would enable us to
analyze the physical content of Fock coherent states \textit{from the
non-perturbative perspective}.  This will provided by the notion of
shadow states.

For concreteness, let us focus on the vacuum $<\!  V_F^\r\!\mid $.  It
follows from the expression (\ref{vac}) that, given \textit{any} flux
network function $\N_{\beta, \n}$ associated with a fixed graph
$\beta$, the action of $<\! V_F^\r\!\mid $ on $\N_{\beta, \n}$ can be
written as:
\be 
<\!  V_F^\r\!\mid \, \N_{\beta, \n} \!  > \,\, = \,\,
e^{ -\frac{\hbar}{2} \sum_{IJ} G_{IJ}n_I n_J }\,\,  
:= \,\, \int_{\A_\beta}  d\mu_\beta^{o}\, \bar{V}_\beta^\r\, 
\N_{\beta, \n}
\ee
where $I, J$ now label the edges of the graph $\beta$ and where the 
function $V_\beta^\r(\hb_\beta)$ on $\A_\beta$ is given by
\be \label{shadow}
V_\beta^\r(\hb_\beta) = \sum_{\n}\, e^{ -\frac{\hbar}{2} \sum_{IJ} 
G_{IJ}\, n_I n_J} \,\, \N_{\beta, \n} (\hb_\beta)
\ee
This implies that the action of the element $<\!  V_F^\r\!\mid $ of
$\cyl^\star$ on \textit{any} element $\varphi_\beta$ of $\cyl$ (based on
$\beta$) is the same as the inner product of $V_\beta^\r(\hb)$ with
$\varphi_\beta$ (on $L^2(\A_\beta, d\mu^\r_\beta)$).  Therefore, we will
refer to the function $V_\beta^\r$ on $\A_\beta$ as the \textit{shadow
of the Fock vacuum} $<\!  V_F^\r\!\mid $ on the graph $\beta$.  The
set of shadows on all possible graphs captures the full information in
$<\!  V_F^\r\!\mid $.  Furthermore, since each shadow is an element of
a \textit{Hilbert space} $L^2 (\A_\beta, d\mu^\r_\beta)$ associated
with \textit{some} graph $\beta$, we can now take expectation values
and and calculate fluctuations of operators in the shadow states.  The
construction can be repeated for any Fock coherent state.  Therefore,
using shadow states we can spell out the precise sense in which Fock
coherent states are semi-classical also from the non-perturbative
perspective and to understand limitations of the Fock description in
the ultraviolet regime \cite{afl}.
\bigskip

\textit{Heat kernels and quantum gravity}
\medskip

Let us now return to non-perturbative gravity and indicate how our
$U(1)$ constructions can be extended to that case.  Our goal is to
find an element $<\!\M^\r\!\mid$ of gravitational $\cyl^\star$ which
represents a semi-classical state peaked at the trivial Minkowskian
initial data (constant triads ${}^0\!E^a_i$ and zero gravitational
connection, $A_a^i= 0$).  Now the gauge group is $SU(2)$ and flux
network states are replaced by the spin network states $\N_{\alpha,
\vec{j}, \vec{I}}(\hb) $, where half-integers $j_e$ (or, irreducible
unitary representations of $SU(2)$) label edges $e$ of the graph, and
invariant operators $I_v$ on (the tensor product of `incoming and
outgoing' representations of) $SU(2)$ label its vertices $v$
\cite{rs2,jb2,jb3,almmt}.  In spite of the fact that the situation is
now technically more complicated, every element of $\cyl^\star$ can be
still expanded in terms of the dual spin network basis, $<\!
\N_{\alpha, \vec{j}, \vec{I}}\!\mid$.  Our goal is to find the
expansion coefficients of $<\!\M^\r \!\mid$ in this basis.

To carry out this task let us first reformulate the construction of
the shadows of the $\U(1)$ Fock vacuum using `heat kernel' ideas
\cite{al3,almmt2} which can be then extended directly to the
non-Abelian context. Let us begin by introducing a family of
operators. Fix a vertex $v$ of a graph $\beta$ and label by $I$ the
edges which have $v$ as an endpoint. Define operators $X_I$ via their
action on any cylindrical function $\varphi_\beta$ associated with
$\beta$:
\be\label{X} 
(X_I\, \varphi_\beta) = \cases{
-\hb(e_I)\, {\partial \varphi_\beta / \partial (\hb(e_I))}\, ,
& if $v$ is the target of $e_I$,  \cr
\hb(e_I)\, {\partial \varphi_\beta / \partial (\hb (e_I))}\, ,
& if $v$ is the source of  $e_I$,\cr}
\ee
where on the right side we have regarded $\varphi_\beta$ as a function
of the $N$ holonomies $\hb(e_1), \ldots ,\hb(e_N)$. Thus, if $v$ is
the target, $X_I$ is the right invariant vector field on the copy of
$U(1)$ associated with the $I$th edge and, if it is the source, the
left invariant vector field. (While the distinction between right and
left invariant vector field is superfluous for $U(1)$, it becomes
important in the $SU(2)$ case discussed below.)  Using the fact that
the flux networks are eigenstates of $X_I$, the shadow vacuum
$V_\beta^\r$ of (\ref{shadow}) can be expressed as:
\be \label{shadow2}
V_\beta^\r(\hb_\beta) = \sum_{\n}\, e^{ \frac{\hbar}{8} \sum_{v,v'} 
\sum_{I,I'}\, G_{II'}\, X_I X_{I'}} \,\, \N_{\beta, \n} (\hb_\beta)
\ee
where, $I$ and $I'$ now label edges associated with vertices $v$ and
$v'$ respectively. The exponent is a negative definite self-adjoint
operator on $L^2(\A_\beta, d\mu^{o}_\beta)$, reminiscent of heat
kernels.%
\footnote{If $G_{II'}$ were $\delta_{II'}$, the exponent would simply
be the product of Laplacians, one for the copy of the group associated
with each edge and $V_\beta^\r$ would be a group coherent state on
$[U(1)]^N$ a la Hall \cite{bh}.}
Indeed, by using the Peter-Weyl expansion of the Dirac $\delta$
distribution on $\A_\beta \simeq [U(1)]^N$, one can re-express
(\ref{shadow2}) more conveniently as
\be \label{shadow3}
V_\beta^\r(\hb_\beta) = e^{ \frac{\hbar}{8} \sum_{v,v'} 
\sum_{I,I'} G_{II'}X_I X_{I'}} \,\, \delta_0(\hb_\beta)   
\ee
where $\delta_0(\hb_\beta)$ is the Dirac-distribution on $(\A_\beta,
d\mu_\beta^{o})$ peaked at the zero connection. On any fixed graph
$\beta$, the right side is a continuous function on
$\A_\beta$. However, as remarked earlier, the family of functions
obtained by varying graphs fails to be consistent, i.e., does not
provide a single $\mu_0$-mesurable function on $\Ab$.  On the other
hand, one can show that the family of operators on the right side of
(\ref{shadow3}) \textit{is} consistent whence the right side provides
a consistent family of \textit{distributions}. Hence, the element of
$\cyl^\star$ representing the Fock vacuum can be written as
\be
<\! V_F^\r\!\mid \varphi \! > \,\, =\,\,  \int_{\Ab}  d\mu_o\,\, 
( e^\Theta \,\, \delta_o(\hb ))\,\, \varphi \qquad\qquad
\forall \varphi
\in \cyl
\ee 
where $\Theta$ is the projective limit of the family of operators
$[\frac{\hbar}{8} \sum_{v,v'} \sum_{I,I'} G_{II'}X_I X_{I'}]$ on
$(\A_\beta, d\mu_\beta^{o})$ and $\delta_0(\hb)$ is the
Dirac-distribution on $(\Ab, d\mu_o )$ peaked at the zero generalized
connection. Similarly, the coherent state (\ref{coh}) can be expressed
as
\be \label{coh2}
<\! \Psi_{(A_o,E_o)}^\r\!\mid\, \varphi\!> \,\, =\,\, 
\int_{\Ab}  d\mu_o\,\, 
( e^\Theta \,\, \delta_{\tilde{f}}(\hb ))\,\, \varphi \qquad \qquad
\forall \varphi \in \cyl
\ee
where $\delta_{\tilde{f}}$ is the Dirac-distribution peaked at
$\tilde{f} = A_o(k) - (i/ \mid\! k \! \mid) E_o(k)$, regarded as as a
(complex) generalized connection.

We will now illustrate how these considerations can be extended to
quantum gravity. To carry out this task, $\U(1)$ has to be replaced by
$\SU(2)$. This is a non-trivial step because of the subtleties
associated with non-Abelian gauge invariance. Specifically, the $U(1)$
operators $X_I$ are now replaced by $SU(2)$ operators $J_I^i$ where
$i$ is an ${\rm{su}}(2)$ Lie-algebra index \cite{al3,al4} and, to
construct the analog of the operator $\Theta$, we must find a way to
contract these ${\rm{su}}(2)$ indices associated with
\textit{different vertices}.  This would have been a major obstacle
without recourse to a background connection to transport the indices
between different vertices.  Fortunately, however, since our task is
to construct semi-classical states peaked at given classical fields
$(A_a^i, E^a_i)$, we can use the given $A_a^i$ as the background
connection. Then, for $<\!  \M^\r \!\mid$, the required parallel
transport is trivial since the background connection vanishes!

Since the state is to be peaked at a constant triad ${}^0\!E^a$, and
since the triad and the connection have \textit{different} physical
dimensions, we need to introduce a new length scale%
\footnote{In the Maxwell theory, this can be seen by using a natural
regularization technique to construct a generalized coherent state peaked 
at zero connection and a constant electric field \cite{afl}.}
$\ell$ (to be fixed by the Planck length $\lp$ and the macroscopic
scale $L$ of physical interest; see below).  Our candidate shadow
states $<\!  \M^\r_\beta \!\mid$ are thus given by
\be \label{shadow4} 
\M^\r_\beta (\hb_\beta) = 
e^{\frac{\hbar}{8}\, \sum_{v,v'} \sum_{I,I'} G_{II'}\, K_{i,i'}\, 
J_I^i J^{i'}_{I'}} \,\, \delta_{{}^0\!E}(\hb_\beta)\, .
 \ee
Here $\delta_{{}^0\!E}(\hb)$ is the Dirac-distribution peaked at
$-(i/\ell) \, {}^0\! E_a^i $ (regarded as a connection on $M$ in the
gauge specified by the background fields under consideration),
$K_{i,i'}$ is the Cartan-Killing metric on ${\rm su}(2)$ (and the
internal indices are transported between vertices $v$ and $v'$ by the
trivial connection). Again, as the graph varies, operators on the
right hand side yield a consistent family. Hence, the right side of
(\ref{shadow4}) constitutes a consistent family of distributions on
the configuration spaces $(\A_\beta, d\mu_\beta^o)$ and thus defines
an element $<\! \M^\r\!\mid$ of $\cyl^\star$. Since our specification
of the background field ($A_a^i = 0$) is not gauge invariant, neither
is the state $<\! \M^\r \!\mid$. However, if we wish, we can easily
group average it over the group $\G$ of (arbitrarily discontinuous)
gauge transformations and obtain a gauge invariant state in
$\cyl^\star$ \cite{almmt}.
 
Properties of this candidate semi-classical state and its
generalizations for other background classical gravitational fields
$(A_a^i, E^a_i)$ are being investigated.
\bigskip

\textit{Statistical geometry}
\medskip

In practice ---particularly for numerical semi-classical calculations
which were recently launched--- it is awkward to have to deal with
\textit{all} graphs. Furthermore it is clear that, to probe a
semi-classical state effectively, the graph should be sufficiently
fine; otherwise the shadow state would be too crude an approximation.
For semi-classical purposes, then, can one restrict oneself to a
judiciously chosen sub-family which is small enough to be manageable
and yet large enough for the associated shadows to capture all the
relevant information contained in the semi-classical state in
$\cyl^\star$ from which they originate? The answer is in the
affirmative.  We will summarize this strategy from the perspective of
quantum gravity, coupled to Maxwell theory; for details, see
\cite{ab}.

Given a 3-manifold $M$ with a positive-definite metric $q^{o}_{ab}$,
using well established techniques from statistical geometry, one can
introduce on it a natural family of (Voronoi) graphs \cite{ab}.  For
simplicity, let us suppose $M$ is a 3-torus, $q^{o}_{ab}$ is flat and
endows $M$ with a volume $V$. Consider a random sprinkling of points
in $M$ with a given mean density $\rho$. Then there is a natural
procedure to construct a simplicial complex and a dual cell complex,
and introduce, from the cell complex, a graph $\alpha_{x_1,\ldots
,x_n}$, labelled by the $n= V\rho$ points of the given sprinkling (and
of course ($M, q^{o}_{ab}, \rho))$.  The construction is
\textit{covariant} in the sense that it does not require any
additional inputs. In particular, then, the family of graphs
$\{\alpha_{x_1,\ldots ,x_n}\}$ is preserved by the action of
isometries of $q^{o}_{ab}$ on $M$. For large $n$, almost all vertices
of graphs are four-valent, whence they are especially well-suited for
quantum geometry \cite{al5}. Furthermore, using techniques from
statistical geometry, one can estimate the number of vertices in any
given `sufficiently large' region and the number of intersections of
the graph with any `sufficiently large' surface with slowly varying
extrinsic curvature. These estimates facilitate the task of
calculating expectation values and fluctuations of geometric and
Maxwell operators in candidate shadow states based on these Voronoi
graphs. Thus, this family appears to be large enough to capture
`enough' shadow states and yet small enough to be manageable.

We will conclude by summarizing the qualitative indications that have
been obtained from these and related calculations. First, one loop QED
corrections to the Maxwell vacuum $<\! V_F^\r\!\mid$ in Minkowski
space have been calculated by Dreyer and Ghosh \cite{dg}. Their result
shows that the parameter $r$ is related to the cut-off, tending to
zero as the ultra-violet cut off in the momentum space goes to
infinity. On the other hand, quantum geometry (without Maxwell fields)
has also been examined using Voronoi graphs \cite{ab}. One finds that
no state based on this graph can serve as the shadow of the
semi-classical state in $\cyl^\star$ peaked at $q^{o}_{ab}$, unless
the mean separation $a = 1/\root3\of{\rho}$ between its vertices is
greater than (a certain multiple of) the Planck length
$\lp$\cite{ars2,ab}. Thus, quantum geometry has a built-in cut off at
$\lp$, whence we are led to set $r \ge\lp$. Suppose we are interested
in observables associated with a macroscopic scale $L$. (For example,
we may be interested in measuring areas or fluxes of electric and
magnetic fields across surfaces of a characteristic length greater
than or equal to $L$.) Then, from the `polymer perspective', one finds
that the Fock description is adequate only if $L >> r \ge \lp$.  For
questions involving frequencies comparable to or greater than the
Planck frequency, the Fock description is a poor approximation to the
`fundamental', polymer description. It seems  likely that
sharpened versions of these calculations will lead to detailed answers
to the two questions with which we began.

\bigskip

\textbf{Acknowledgements} We would like to thank Stephen Fairhurst and
Madhavan Varadarajan for stimulating discussions and Christian
Fleischhack, Jos\'e Mour\~{a}o, Jorge Pullin and Jos\'e Velhinho for
clarifying correspondence.  This work was supported in part by the NSF
grants PHY-0090091, INT97-22514, the Polish CSR grant 2 P03B 060 17,
the Albert Einstein Institute and the Eberly research funds of Penn
State.

\end{document}